\documentclass[aps,prb,twocolumn,superscriptaddress]{revtex4-1}

\usepackage{amsfonts}
\usepackage{graphicx}
\usepackage{bm}

\begin{document}
\title{Electrically controlled pinning of Dzyaloshinskii-Moriya domain walls}

\author{Koji Sato}
\affiliation{WPI Advanced Institute for Materials Research, Tohoku University, Sendai 980-8577, Japan}
\author{Oleg A. Tretiakov}
\email{olegt@imr.tohoku.ac.jp}
\affiliation{Institute for Materials Research, Tohoku University, Sendai 980-8577, Japan}
\affiliation{School of Natural Sciences, Far Eastern Federal University, Vladivostok 690950, Russia}

\begin{abstract}
We propose a method to all-electrically control a domain-wall position in a ferromagnetic nanowire with Dzyaloshinskii-Moriya interaction. The strength of this interaction can be controlled by an external electric field, which in turn allows a fine tuning of the pinning potential of a spin-spiral  domain wall. It allows to create more mobile pinning sites and can also be advantageous for ultra-low power electronics.

\end{abstract}

\maketitle

Within the last decade there has been enormous interest in both experiments and theories of the magnetic domain wall (DW) motion.\cite{Yamaguchi04,  Nakatani03, Tatara04, Zhang04, Beach05, KlauiPRL05, Meier07, Moriya08,  Duine07, Tserkovnyak2008, Tretiakov08, Clarke08, BeachPRL09,  Lucassen09, Rhensius10, IlgazPRL10, Singh10, Krivorotov10, Thomas2010,  Min10, Brataas2010, Hertel2010, Tretiakov2012, Thiaville2012, Beach2013, Parkin2013,  Brataas2013, Ohno2014} Controlling DWs in a nanowire has been one of the great challenges for successful applications of spintronic memory, logic, and sensor nanodevices.\cite{Parkin:racetrack08, Allwood01, Allwood02} For the manipulation of magnetic DWs spin currents and local Oersted fields have been used.\cite{IlgazPRL10} Meanwhile, there are certain limitations in applying these schemes: employing a spin polarized charge current suffers from the Joule heating,\cite{Tretiakov:losses} whereas using the Oersted field has difficulty in making scalable systems. Less dissipative and more scalable method of controlling DWs would thus provide meaningful steps for the further development in DW based devices such as race-track memories.\cite{Parkin:racetrack08} 

For such technologies based on manipulation of DWs, controlling the position of a DW is an important task.\cite{Beach2013} Conventionally, mechanical notches have been used to fix the DW position,\cite{HayashiPRL2006, Bogart2009, Munoz2011, YuanPRB2014} which provide pinning potential for DWs by structural means. A potential drawback of this approach is the lack of mobility of the pinning site and difficulty in fabrication. Therefore, it is more desirable to achieve the DW pinning where the position can be easily adjusted along the nanowire, especially by all-electric means. 
 
\begin{figure}[ht]
\includegraphics[width=\linewidth]{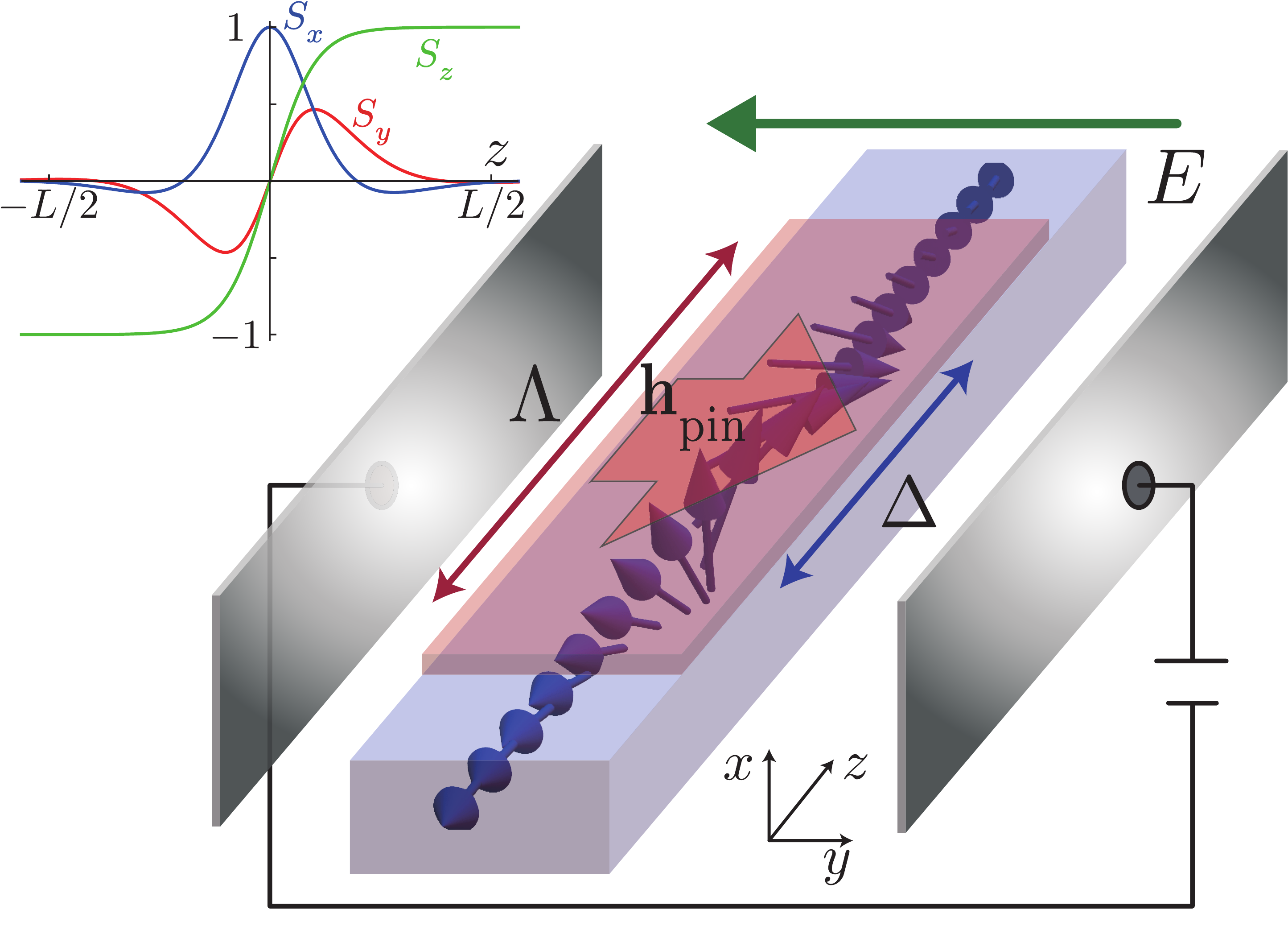}
\caption{The setup for a ferromagnetic nanowire with a domain wall (blue arrows in the center) in proximity to a strong single-domain ferromagnet (red region) under electrostatic gating.  The DW profile has the spiral pitch $\Lambda$ and DW width $\Delta$. The electrostatic gates provide an external electric field $\mathbf E$. The ferromagnet provides the pinning field $\mathbf h_{\rm{pin}}$, which is represented by a large red arrow. The inset in the left top corner shows the magnetization components $S_x$ (blue), $S_y$ (red), and $S_z$ (blue) vs.  the coordinate ($z$) along the wire.}
\label{fig1}
\end{figure}

In this Letter, we propose an approach to control the pinning of a DW in a thin ferromagnetic nanowire by an external electric field, see Fig.~\ref{fig1}. We consider a nanowire with an easy-axis anisotropy along the wire, exchange interaction, and Dzyaloshinskii-Moriya interaction (DMI),\cite{Dzyaloshinskii58, Moriya60} where transverse DWs can be formed when the anisotropy is sufficiently strong. Due to DMI, the domain wall exhibits spin-spiral structure, whose pitch is determined by the DMI strength. As an external electric field is applied, the strength of DMI can be modified, leading to the change in the pitch of the spiral DW. Consequently, the total transverse magnetization of the DW either increases or decreases, and the interaction between the DW and the pinning ferromagnet (or magnetic field) is strengthen or weakened correspondingly, Fig.~\ref{fig1}. Hence, the pinning strength can be controlled by electrically manipulating the strength of DMI. This proposal provides the means to control DW positions all-electrically, which may assist for more precise DW operation. Furthermore, the threshold current for driving a DW can be reduced by this method, which could also be beneficial for low-power spintronics applications.

We consider a thin ferromagnetic nanowire and employ an effective one-dimensional classical spin model with isotropic exchange interaction, uniaxial anisotropy, and DMI. The easy axis of the anisotropy is taken along the wire ($z$-axis), and the DMI vector $\mathbf D$ points in $z$-direction. The Hamiltonian describing such a system is given by
\begin{eqnarray}
\label{hamiltonian a}
\mathcal H &=& \int dz\left[
\frac{J}{2}\left(\partial_z\mathbf S\right)^2+D\mathbf S\cdot\left(\mathbf e_z\times\partial_z\mathbf S\right)-K S_z^2\right]\nonumber\\
&= &\int dz\left\{\frac{J}{2}\left[(\partial_z\theta)^2+\sin^2\theta(\partial_z\phi)^2\right]\right.\nonumber\\
&&+(K-D\partial_z\phi)\sin^2\theta\bigg\}+\rm{const}\,,
\end{eqnarray}
where $\mathbf S=\mathbf M/M=(\sin\theta\cos\phi,\sin\theta\sin\phi,\cos\theta)$ is the unit vector of the magnetization $\mathbf M$, $J$ is the exchange constant, $D$ is the intrinsic DMI constant, and $K$ is the anisotropy constant. The above material parameters are appropriately rescaled from the original ones to describe this effective one-dimensional model.\cite{Tretiakov2010} The effect of the transverse anisotropy is neglected for static pinning in this paper. The $B$20 structure of ferromagnets, such as MnSi, which lacks inversion symmetry, can be a good candidate to model the ferromagnet with DMI.

This model supports two distinct types of solutions depending on the relative strength of $J$, $D$, and $K$. The static solution for the magnetic texture $\mathbf S(z)$ can be found by employing a variational method, i.e. seeking for the energy minimum. By varying $\theta$ and $\phi$, we find the following equations: 
\begin{eqnarray}
\label{variation1}
&&J\partial_z^2\theta-\left[\frac{J}{2} (\partial_z\phi)^2  +K-D\partial_z\phi\right]\sin2\theta=0\,,\\ \label{variation2}
&&\partial_z\left[\left(J\partial_z\phi-D\right)\sin^2\theta\right]=0\,.
\end{eqnarray}
In the regime of sufficiently strong DMI relative to both $J$ and $K$ (more precisely $2JK<D^2$), the energy is lowered by making a spiral texture due to DMI, which can overcome the energy increase from varying texture and deviation from $S_z=1$. The solution in this regime is $\mathbf S=(\cos(z/\Lambda+\phi_0),\sin(z/\Lambda+\phi_0),0)$, which is a spiral structure rotating around $z$ axis, with $\Lambda=J/D$ being the pitch of the spiral, and $\phi_0$ being an arbitrary phase. 

In the other regime $2JK>D^2$, where the anisotropy and exchange interaction are sufficiently strong, the entire texture favors to align along $z$ axis. This configuration has the lowest energy because it reduces both the anisotropy and exchange interaction. Since Eqs.~(\ref{variation1}) and (\ref{variation2}) allow us to determine not only the global minimum of the energy landscape but also the local minima, DW solutions can be found by appropriately setting the boundary conditions. Solving Eqs.~(\ref{variation1}) and~(\ref{variation2}) with the boundary conditions for a tail-to-tail DW, i.e. $S_z(\pm\infty)=\pm 1$, we find $\theta(z)=\cos^{-1}\tanh((z-z_0)/\Delta)$ and $\phi=(z-z_0)/\Lambda+\phi_0$, where $z_0$ is the location of the DW center and $\phi_0$ is an arbitrary phase. Here $\Delta$ is the width of the DW, which is given by $\Delta^{-2}=\Delta_0^{-2}-\Lambda^{-2}$, where $\Delta_0=\sqrt{J/2K}$ corresponds to the DW width without DMI. The coordinate dependence of a spin texture is obtained by plugging $\theta(z)$ and $\phi(z)$ into $\mathbf S(z)=(S_x,S_y,S_z)=(\sin\theta\cos\phi,\sin\theta\sin\phi,\cos\theta)$. Note that the DW width $\Delta=J/\sqrt{2JK-D^2}$ depends on the strength of DMI. Within the DW width, the texture deviates significantly from $S_z=\pm1$, and the energy can be lowered by making a spin-spiral structure on top of the DW structure when DMI is present. Considering the interplay of the DMI, exchange, and anisotropy, as DMI becomes stronger, wider DMs are energetically favored.

In the present analysis, we are interested in manipulating DWs by an electric field, thus hereafter we consider the regime where DWs can be formed, i.e. $2JK>D^2$ (or $\Lambda > \Delta_0$). Because the DW width depends on the DMI strength, in the following we present a way to control the DMI strength by an external electric field, and thus in turn the DW width. 

In a lattice model, the DMI between two spin sites $i$ and $j$ is given by $\mathbf D_{ij}\cdot(\mathbf S_i\times\mathbf S_j)$, where $\mathbf D_{ij}$ is the DMI vector, and $\mathbf S_{i(j)}$ is the spin at the site $i(j)$.\cite{Dzyaloshinskii58, Moriya60} The direction of the DMI vector is governed by the crystalline symmetry. The DMI in the Hamiltonian of Eq.~(\ref{hamiltonian a}) is obtained as the continuous limit of the above discrete form of the DMI. As an example of such a nanowire, we suppose that a series of layers are stacked to constitute the nanowire. 

In the present model, the crystalline symmetry of the wire has been appropriately chosen so that the DMI vector points in $z$-direction, namely the DMI vector associated with each layer in the nanowire points in $z$-direction. Each of these layers has a number of spins and associated DMI vectors, and the DMI strength parameter $D$ appearing in Eq.~(\ref{hamiltonian a}) is the total value of the DMI vectors of a layer with infinitesimal thickness $dz$. Crystal symmetry can be manipulated by applying an external electric field $\mathbf E$, and as a result the DMI also changes. For now, let us consider one of the layers constituting the nanowire. The magnetoelectric response of DMI has been studied in the case of two-dimensional system.\cite{siratoriJPSJ80, *katsuraPRL05} When an external electric field is applied in the plane of this layer ($xy$-plane), the DMI vector $\mathbf D_{ij}$ is modified by 
\begin{equation}\label{D}
\delta\mathbf D_{ij}\propto\hat{\mathbf e}_{ij}\times\mathbf E\,,
\end{equation}
where $\delta\mathbf D_{ij}$ is the change of the DMI vector, $\mathbf E$ is the electric field applied in $xy$-plane, and $\mathbf e_{ij}$ is the unit vector connecting the sites $i$ and $j$ in the layer, which also lies in $xy$-plane. As a consequence, the change of the DMI vector points in $z$-direction, i.e. $\delta\mathbf D=\delta D\mathbf e_z$ where $\delta D$ can be regarded as the total change of the DMI vector in the layer. Since the original DMI vector points in $z$-direction, we conclude that the DMI strength under the electric field is given by $\tilde D=D+\delta D$. Thus, we find that the strength of DMI can be modified by a transverse electric field, and as a consequence, the pitch of the spiral and the DW width are modified as $\tilde\Lambda=J/\tilde D$ and $\tilde\Delta^{-2}=\Delta_0^{-2}-\tilde\Lambda^{-2}$, respectively. In order to implement this effect, we propose a setup shown in Fig.~\ref{fig1}, where the nanowire is placed between the planar gates to apply an external electric field transverse to the wire. This setting allows us to control the DW width and spiral pitch by gating. \footnote{The effect of an external electric field can be not only on the DMI but also on the perpendicular anisotropy, \cite{maruyama2009} which could introduce a knob to control the structure of DWs and thus in turn the pinning strength. However, the anisotropy considered in our model is dominated by the dipolar interactions leading to a shape anisotropy, and it is only weakly modified by an external electric field unlike the perpendicular anisotropy. We thus neglect the change in the anisotropy in the present analysis.} 

Our analysis takes a single-domain nanowire as a starting point, so the thickness and width of the wire are limited by the single-domain condition. We estimate the thickness and width of such nanowire to be roughly $\sim$10 nm and $\sim$200 nm, respectively, for materials such as Py.\cite{Parkin:racetrack08} DW width in such a nanowire is of the order of the wire's width. As for the gate, the lateral dimensions of the plates are expected to be larger than the DW width ($\sim$500 nm), and the distance between the plates is of the same order as the wire thickness ($\sim$10 nm).  Furthermore, the strength of the electric field provided for gating should be around 10-100kV/cm to observe changes in DMI.\cite{rovillain2010, zhang2014,nawaoka2015} Assuming the plates separation is $\sim$10 nm, $\sim$0.1 V can provide $\sim$100 kV/cm.

In general, the pinning field can be created by either structural feature of the wire such as a notch or an external magnetic field. Below we explain how the changes in the DMI vector and DW width can be utilized to control the pinning of DWs, which is distinct from the aforementioned methods. We describe the pinning of a DW by the interaction between the average magnetization of the DW and an effective pinning field.

In the current proposal, we consider a system of a nanowire supporting a DW in proximity to a single domain ferromagnet, whose stray field serves as the effective pinning field for the DW, see Fig.~\ref{fig1}. We consider the pinning field pointing in the transverse direction with respect to the wire axis. Such pinning by stray magnetostatic fields of a magnet has been recently experimentally studied,\cite{OBrien2011} and furthermore this experiment shows that the structure of the DW is relatively unperturbed.

The strength of the DW pinning depends on the interaction between this pinning field and the average magnetization of the DW, thus the larger the average value of the DW magnetization in the transverse direction is, the stronger the pinning is. We model the pinning of the DW by Zeeman type of interaction between the average transverse component of the DW magnetization $\mathbf S^{\rm{ave}}=(S^{\rm{ave}}_x,S^{\rm{ave}}_y,0)$ and the pinning field $\mathbf h_{\rm{pin}}$ from the proximity ferromagnet. The energy associated with the pinning is $\mathcal H_{\rm{pin}}\propto -\mathbf S^{\rm{ave}}\cdot\mathbf h_{\rm{pin}}$. As we show below, $\mathbf S^{\rm{ave}}$ depends on the pitch of the spiral $\tilde\Lambda$. Since the strength of the pinning is directly proportional to $\mathbf S^{\rm{ave}}$, and it can be controlled by changing $\tilde\Lambda$ via electrostatic means. 

Suppose the length of the proximity ferromagnet and gated region is $L$ (extending in $-L/2\leq z\leq L/2$), which is much larger than the DW width ($L\gg\tilde\Delta$), so that the entire nanowire system can be treated with the modified DMI constant $\tilde D$. Note that the transverse components of the magnetization are given by $(S_x,S_y,0)=\sin\theta(\cos\phi,\sin\phi,0)$, where $\sin\theta(z)=1/\cosh(z/\tilde\Delta)$ and $\phi(z)=z/\tilde\Lambda+\phi_0$, i.e. the DW is centered at $z=0$. Under this condition, their average values can be estimated as
\begin{eqnarray}
\label{averageS}
\mathbf S^{\rm{ave}}&=& \int^{L/2}_{-L/2} \frac{dz}{L}\frac{1}{\cosh\left(z/\tilde\Delta\right)}(\cos\phi(z),\sin\phi(z),0)\nonumber\\
&=& S^{\rm{ave}}(\cos\phi_0,\sin\phi_0,0)\,,
\end{eqnarray}
where the magnitude of $\mathbf S^{\rm{ave}}$ is
\begin{equation}\label{Save}
S^{\rm{ave}}
\approx \frac{\pi\tilde\Delta}{L\cosh\left(\pi\tilde\Delta/2\tilde\Lambda\right)}\,,
\end{equation}
which is exact for $L\to \infty$. As the effect of $\bm H_{\rm pin}$ is considered, the arbitrary angle $\phi_0$ in Eq. (5) is determined in such a way that $\bm S^{\rm ave}$ points in the direction of $\bm H_{\rm pin}$.

\begin{figure}[htb]
\includegraphics[width=\linewidth]{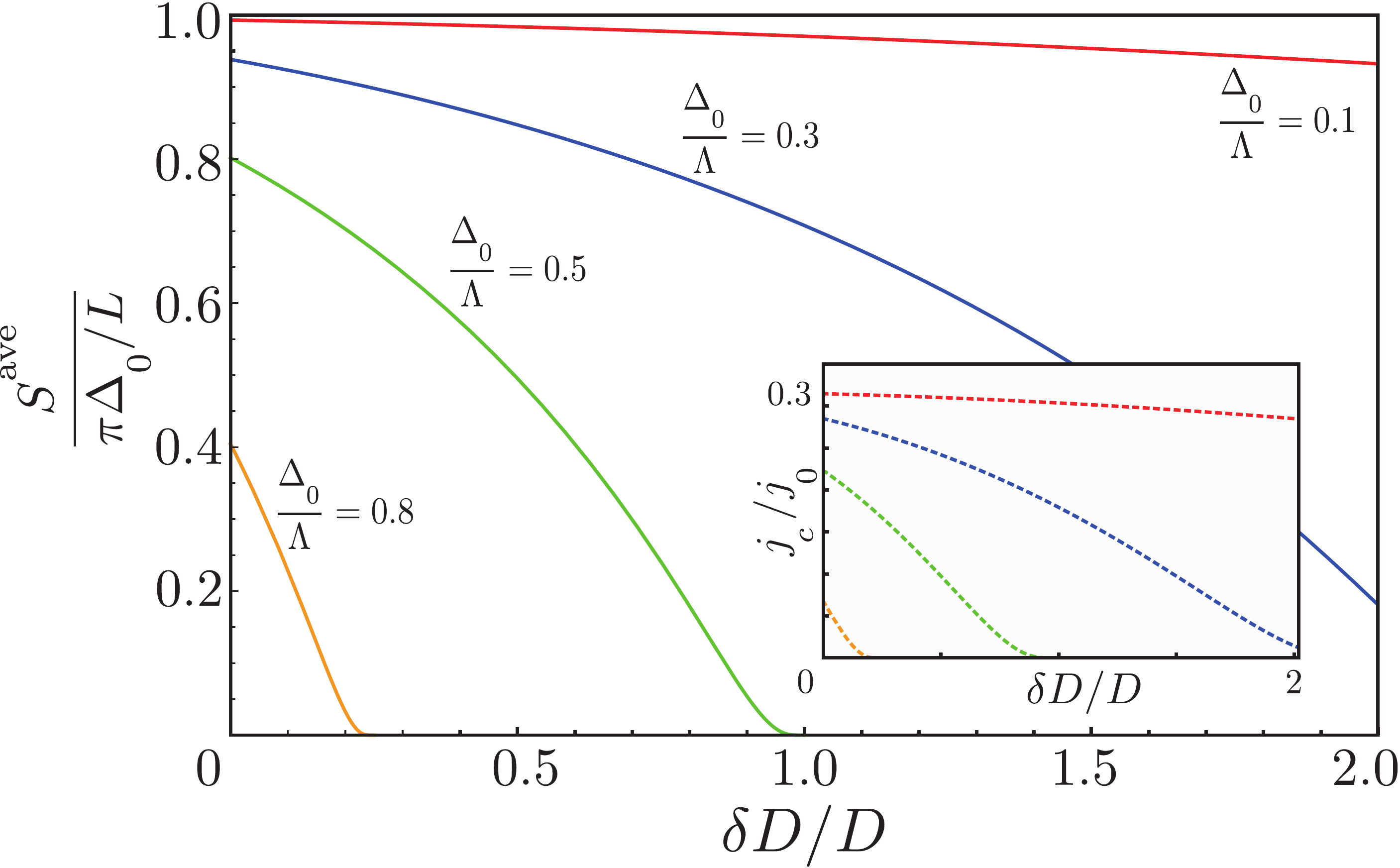}
\caption{The magnitude of the average magnetization of the domain wall, $S^{\rm{ave}}$, as a function of the change in the DMI constant $\delta D$ due to the external electric field. The maximum value of the average field depends on the ratio of the DW width without the DMI, $\Delta_0$, and the pitch of the spiral structure, $\Lambda$. The figure shows the average magnetization for various values of $\Delta_0/\Lambda$. The inset depicts the critical current $j_c$ vs. $\delta D$. The color of the dotted lines for $j_c$ plot corresponds to the same color for $S^{\rm{ave}}$ plot with corresponding values of $\Delta_0/\Lambda$.}
\label{fig2}
\end{figure}

Since $\tilde\Delta$ and $\tilde\Lambda$ depend on $\delta D$, we can plot $S^{\rm{ave}}/(\pi\Delta_0/L)$ as a function of $\delta D/D$, as shown in Fig.~\ref{fig2} for various values of the ratio between the DW width and the spiral pitch, $\Delta_0/\Lambda$. Note that the regime where a DW can be formed is $\Lambda>\Delta_0$, thus the ratio is $\Delta_0/\Lambda<1$. For all values of $\Delta_0/\Lambda$, $S^{\rm{ave}}$ is a monotonically decreasing function as $\delta D$ is increased. The length of the spiral pitch is decreased by increasing $\delta D$, which means the spiral tends to rotate more within a given DW width. Since the transverse components of the magnetization spiral more, the average values of the transverse components within the DW are reduced. 

Furthermore, we can observe the tendency that the smaller the ratio $\Delta_0/\Lambda$ is, the less it is affected by the change of the DMI, see the red curve in Fig.~\ref{fig2}. When $\Delta_0/\Lambda$ is small, the texture hardly spirals over the DW region since the spiral pitch is much longer than the DW width. Since the transverse magnetization does not change much over the DW, $S^{\rm{ave}}$ becomes large. In this regime, a small change of the spiral pitch does not affect the situation where the texture hardly spiral over the DW, hence $S^{\rm{ave}}$ cannot change substantially. On the other hand, when $\Delta_0/\Lambda$ is near 1, the spiral pitch is about the same as (but a little longer than) the DW width. As the spiral pitch shortens to approach the DW width, $S^{\rm{ave}}$ quickly approaches zero. From Eq.~(\ref{Save}), we can see the transition between these two extreme situations ($\Delta_0/\Lambda\rightarrow 0$ and $1$). In fact, when $\Delta_0/\Lambda$ approaches 1, the average magnetization $S^{\rm{ave}}$ becomes exponentially sensitive to the change in the DMI strength, which may be an interesting regime for application purposes. 

Although we have so far focused only on the static pinning effect, the dynamics of the magnetic texture can also play an important role in controlling the DW pinning. As a spin-polarized current runs through the nanowire, a spin transfer torque is applied to the magnetization texture. The dynamics of the magnetic texture under these circumstances is described by Landau-Lifshitz-Gilbert equation with the spin transfer torque.\cite{Tatara04, Zhang04, Thiaville05} Employing this equation for a DW in a nanowire with a small transverse anisotropy, one can find the critical current $j_c$, \cite{Tretiakov2010} 
\begin{equation}
\label{j_c}
j_c=\frac{\pi\alpha K_\perp}{|\alpha-\beta|}\frac{{\Delta}^2/\Lambda}{\sinh(\pi\Delta/\Lambda)}\,,
\end{equation}
above which the magnetization in the DW starts to freely rotate around the axis of the nanowire, leading to the dynamic depinning of the DW. Here $K_\perp$ is the transverse anisotropy constant, $\alpha$ is the Gilbert damping constant,\cite{Gilbert04} and $\beta$ is the nonadiabatic spin transfer torque constant.\cite{Tatara04, Thiaville05} When $j>j_c$, the DW texture rotates as it propagates. On the other hand, when $j<j_c$ the DW propagates without rotating. It is thus easier to depin a DW in the regime $j>j_c$, since the texture rotates around to mask the effect of the pinning potential. Therefore, as far as depinning is concerned, it is preferred to operate with lower $j_c$, so that a small amount of current $j$ can depin the DW.    Remarkably, the current needed for the domain-wall depining in the proposed scheme can be much smaller than one used in conventional methods with notches because of the exponential factor in Eq.~(\ref{j_c}), therefore it is advantageous for ultralow-power electronics.

In order to incorporate the effect of the electric gating, we replace $\Lambda\rightarrow\tilde\Lambda$ and $\Delta\rightarrow\tilde\Delta$. The critical current $j_c$ is also a monotonically decreasing function with respect to $\delta D$ as seen in the inset of Fig.~\ref{fig2}. The inset shows $j_c/j_0$ in terms of $\delta D/D$, where $j_0=\pi K_\perp\alpha\Delta_0/|\alpha-\beta|$. Therefore, we can reduce the critical current by increasing $\delta D$, which can also assist the process of the DW depinning. 
  
To summarize, we have proposed a setup to pin and depin spiral domain walls by all-electric means though the DMI control via external electric fields. This method allows to control the DW pinning indirectly, without  electrically affecting the magnetization.\cite{Cherifi2014} It may simplify race-track nanowire fabrication and give flexibility in moving around the pinning sites.

The authors thank G.S.D. Beach and M. Kl\"{a}ui for discussions. O.A.T. acknowledges support by the Grants-in-Aid for Scientific Research (Nos. 25800184, 25247056, and 15H01009) from the Ministry of Education, Culture, Sports, Science and Technology (MEXT) of Japan and SpinNet. K. S. is supported by World Premier International Research Center Initiative (WPI), MEXT, Japan and by JSPS KAKENHI Grant No. 15K13531.

\bibliography{ref_electric_control_pinning}

\end{document}